\documentclass[conference]{IEEEtran}
\IEEEoverridecommandlockouts
\usepackage{geometry}
\usepackage{algorithm}
\usepackage{algorithmic}
\usepackage{cite}

\usepackage{amsmath,amssymb,amsfonts}
\usepackage{algorithmic}
\usepackage{graphicx}
\usepackage{textcomp}
\usepackage{xcolor}

\usepackage{bm}
\def\BibTeX{{\rm B\kern-.05em{\sc i\kern-.025em b}\kern-.08em
    T\kern-.1667em\lower.7ex\hbox{E}\kern-.125emX}}
\begin{document}

\title{Over-the-Air Computation in  Cell-Free \\ Massive MIMO Systems 
}

\author{
\IEEEauthorblockN{
Chen~Chen,
Emil Björnson,
Carlo Fischione
}

\IEEEauthorblockA{
Department of Computer Science, KTH Royal Institute of Technology, Stockholm, Sweden\\
Emails: chch2@kth.se, emilbjo@kth.se, carlofi@kth.se \\
}

}

\maketitle

\begin{abstract}
Over-the-air computation (AirComp) is considered as a communication-efficient solution for data aggregation and distributed learning by exploiting the superposition properties of wireless multi-access channels. However, AirComp is significantly affected by the uneven signal attenuation experienced by different wireless devices. Recently, Cell-free Massive MIMO (mMIMO)
has emerged as a promising technology to provide uniform coverage and high rates  by joint coherent transmission.
In this paper, we investigate AirComp in Cell-free mMIMO systems, taking into account spatially correlated fading and channel estimation errors. 
In particular, we propose optimal designs of transmit coefficients and receive combing at different levels of cooperation among access points. 
Numerical results demonstrate that Cell-free mMIMO using fully centralized processing significantly outperforms conventional Cellular  mMIMO with regard to the mean squared error (MSE). Moreover, we show that Cell-free mMIMO using local processing and large-scale fading decoding can achieve a lower MSE than Cellular  mMIMO when the wireless devices have limited power budgets.
\end{abstract}


\section{Introduction}
Future Internet-of-Things (IoT) are expected to aggregate enormous amounts of data or computation results transmitted from a large number of edge devices~\cite{hellstrom2022wireless, yang2020federated}. Over-the-air computation (AirComp) is a resource-efficient method that leverages the waveform superposition property of the multi-access channel to realize fast wireless data aggregation~\cite{csahin2023survey, chen2023over}. 
Different from conventional digital transmissions,
AirComp is an analog communication scheme that enables a large number of wireless devices to communicate in an interference-free manner using the same time-frequency resource. In other words, the  communication resources required by AirComp do not scale with the number of wireless devices, which is in sharp contrast with conventional digital transmissions using orthogonal wireless resource allocation.

Instead of decoding individual signals transmitted by the wireless devices, AirComp aims to reconstruct a function of the transmitted signals. Theoretically, AirComp can reconstruct any nomographic function. The performance of AirComp is measured by the distortion between the reconstructed function and the desired function, which is quantified by the mean squared error (MSE). There have been research efforts devoted to minimizing the MSE~\cite{cao2020optimized, hellstrom2023federated, wen2019reduced, jing2023transceiver}. 
As for single-antenna cases, the authors in \cite{cao2020optimized} jointly optimized the transmit powers at the wireless devices and the denoising factor  at the receiver.
The work in \cite{hellstrom2023federated} introduced a retransmission mechanism and optimized the power control over multiple AirComp transmissions.
  Multiple-input multiple-output (MIMO) has been a key technology in today's 5G communication systems~\cite{bjornson2024towards}.
The work in \cite{wen2019reduced} considered AirComp in a MIMO system and employed zero-forcing based transmit beamforming.
In \cite{jing2023transceiver}, the authors  developed the optimal 
transceiver beamforming design for multi-antenna transmitters and receivers. The main challenge facing AirComp is the computation errors caused by uneven signal attenuation.
However, this challenge
cannot be well addressed by conventional Cellular network architectures.

Recently, Cell-free Massive MIMO (mMIMO) has shown great potential to provide ubiquitous and uniform coverage with high spectral and energy efficiency~\cite{ammar2021user, 7827017, buzzi2019user}. The idea is to densely deploy a large number of access points (APs) that cooperatively serve all the users in the network; this in turn, eliminates cell boundaries \cite{bjornson2019making}.
Albeit Cell-free mMIMO has been widely investigated in terms of communication performance, the AirComp performance has not been well evaluated.
In \cite{sifaou2023over}, the authors studied over-the-air federated learning in scalable Cell-free mMIMO networks. However, the receiver simply adopted suboptimal maximum ratio combining and the wireless devices transmitted at full power.
Hence, how to optimally implement AirComp in Cell-free mMIMO remains largely an open question.

In this paper, we investigate the advantages of implementing AirComp over Cell-free mMIMO networks. 
The major contributions of our work are summarized as follows: 
\begin{itemize}
	\item We develop optimal designs of AirComp under different cooperation levels among APs. More specifically, we jointly optimize the transmit coefficients and  receive combing  considering spatially correlated fading and
channel estimation errors. We provide asymptotic analysis when the transmit powers of the wireless devices approach infinity, which is validated by simulation results.
	\item We numerically evaluate the MSE performance  of the proposed AirComp designs and show how Cell-free mMIMO should be operated to outperform conventional Cellular mMIMO.  
\end{itemize}

Notations: In this paper, scalars, vectors, and matrices are represented by italic letters, boldface lower-case letters, and boldface uppercase letters, respectively. $\mathbf{v}^{T}$, $\mathbf{v}^{H}$, and $\mathbf{v}^{*}$ denote the transpose, conjugate transpose,
and conjugate of a vector $\mathbf{v}$, respectively. 
$\|\mathbf{v}\|$ denotes the $\ell_{2}$ norm of vector $\mathbf{v}$.
$\mathbf{I}_{N}$ represents a $N\times N$ identity matrix and  $\mathbb{C}^{M\times N}$  represents the space of a complex-valued matrix. 
$\mathbb{E}\{\cdot\}$ is
the statistical expectation and $\text{Tr}(\cdot)$ is the trace operation.
$\mathcal{N_{\mathbb{C}}}\left(\mathbf{0}, \mathbf{R}\right)$ denotes the distribution of a multivariate circularly symmetric complex Gaussian variable with zero mean and covariance matrix $\mathbf{R}$. 

\section{System Model}

We consider AirComp in a Cell-free mMIMO system consisting of $L$ 
multi-antenna APs, each equipped with $N$ antennas, and $K$ single-antenna wireless devices, operating in time-division duplex (TDD) mode. 
All APs cooperate to serve the wireless devices via high-speed fronthaul links connected to a central processing unit (CPU).

The wireless channel between AP $l$ and wireless device $k$ is denoted by $\mathbf{h}_{kl}\in \mathbb{C}^{N}$. Under the assumption of spatially correlated Rayleigh fading channels,
we have 
\begin{align}
\mathbf{h}_{kl}\sim \mathcal{N}_{\mathbb{C}}\left(\mathbf{0}, \mathbf{R}_{kl}\right),
\end{align}
where $\mathbf{R}_{kl}\in \mathbb{C}^{N\times N}$ is the spatial correlation matrix and $\beta_{kl}=\text{Tr}\left(\mathbf{R}_{kl}\right)/N$ is the local-average channel gain subject to path loss and shadowing. 

\subsection{Channel Estimation}
We adopt the block fading channel model where the 
time-frequency resources are divided into multiple coherence blocks, each with $\tau_c$ time-frequency samples. The channels remain static and frequency-flat within a coherence block but may change over different coherence blocks.
In each coherence block, $\tau_{p} \left(\tau_{p}\le K\right)$ time-frequency samples are used for channel estimation and the remaining samples are used for signal transmission. 
Specifically, each wireless device is randomly assigned a pilot signal selected from $\tau_{p}$ mutually orthogonal pilot signals. 
Let $\bm{\phi}_{k}\in \mathbb{C}^{\tau_p}$ denote the pilot signal used by wireless device $k$, and $\mathcal{P}_{k}$ denote the set of wireless devices using the same pilot signal as wireless device $k$.
We have $\bm{\phi}_{k}^{H}\bm{\phi}_{i}=0$,  $\forall i\notin \mathcal{P}_{k}$, and $\|\bm{\phi}_{k}\|^{2}=\tau_p$. 
During the channel estimation phase, the received signal at AP $l$ is given by
\begin{align}
\mathbf{Y}_{l}^{\text{pilot}}=\sum_{i=1}^{K}\sqrt{p_{i}}\mathbf{h}_{il}\bm{\phi}_{i}^{H} + \mathbf{N}_{l},
\end{align}
where $p_{i}$ is the pilot transmit power of wireless device $i$ and $\mathbf{N}_{l}\in \mathbb{C}^{N\times \tau_p}$ is the noise matrix with i.i.d. $\mathcal{N}_{\mathbb{C}}\left(0, \delta^2\right)$ entries.
Then AP $l$ correlates the received signal with the normalized pilot signal of wireless device $k$,
$\bm{\phi}_{k}/\sqrt{\tau_{p}}$, and obtains
\begin{align}
\mathbf{y}_{kl}^{\text{pilot}}&=\sum_{i=1}^{K}\sqrt{\frac{{p_{i}}}{{\tau_{p}}}}\mathbf{h}_{il}\bm{\phi}_{i}^{H}\bm{\phi}_{k} + \frac{1}{\sqrt{\tau_{p}}}\mathbf{N}_{l}\bm{\phi}_{k} \nonumber \\
&= \sum_{i\in\mathcal{P}_{k}}\sqrt{p_{i}\tau_{p}}\mathbf{h}_{il} + \mathbf{n}_{kl},
\end{align}
where $\mathbf{n}_{kl}= \frac{1}{\sqrt{\tau_{p}}}\mathbf{N}_{l}\bm{\phi}_{k}\sim \mathcal{N}_{\mathbb{C}}\left(\mathbf{0}, \delta^2\mathbf{I}_N\right)$.
The MMSE estimate of $\mathbf{h}_{kl}$ is given by
\begin{align}
\hat{\mathbf{h}}_{kl}=\sqrt{p_{k}\tau_{p}}\mathbf{R}_{kl}\bm{\Xi}_{kl}^{-1}\mathbf{y}_{kl}^{\text{pilot}},
\end{align}
where 
\begin{align}
    \bm{\Xi}_{kl}=\mathbb{E}\left\{\mathbf{y}_{kl}^{\text{pilot}}\left(\mathbf{y}_{kl}^{\text{pilot}}\right)^{H}\right\}=\sum_{i\in\mathcal{P}_{k}}p_{i}\tau_{p}\mathbf{R}_{il} + \delta^2\mathbf{I}_{N}.
\end{align} 
The MMSE estimate $\hat{\mathbf{h}}_{kl}$ is distributed as  $\hat{\mathbf{h}}_{kl}\sim \mathcal{N}_{\mathbb{C}}\left(\mathbf{0}, \mathbf{B}_{kl}\right)$, where
\begin{align}
\mathbf{B}_{kl}=\mathbb{E}\left\{\hat{\mathbf{h}}_{kl}\hat{\mathbf{h}}_{kl}^{H}\right\}=
{p_{k}\tau_{p}}\mathbf{R}_{kl}\bm{\Xi}_{kl}^{-1}\mathbf{R}_{kl}.
\end{align}
The estimate error $\tilde{\mathbf{h}}_{kl}= \mathbf{h}_{kl} - \hat{\mathbf{h}}_{kl}$ is independent of $\hat{\mathbf{h}}_{kl}$  and is distributed as $\tilde{\mathbf{h}}_{kl}\sim \mathcal{N}_{\mathbb{C}}\left(\mathbf{0}, \mathbf{C}_{kl}\right)$, where $\mathbf{C}_{kl} = \mathbf{R}_{kl}-\mathbf{B}_{kl}$.

\section{AirComp Design for Three Levels of Receiver Cooperation}
In this section, we develop optimal AirComp designs under different cooperation levels of APs. 
Let $s_{k}$ denote the transmit signal of wireless device $k$. Without loss of generality, we assume that $s_{k}$'s are i.i.d. random variables with zero mean and unit variance~\cite{cao2020optimized, hellstrom2023federated}. In this paper, we are interested in computing the arithmetic mean function of $s_{k}$'s, i.e., 
\begin{align}
f=\frac{1}{K}\sum_{k=1}^{K}s_k.
\end{align}
The extension to other nomographic functions is straightforward by applying pre- and post-processing functions at the transmitters and receivers, respectively.

The received signal at AP $l$ is given by
\begin{align}
\mathbf{y}_{l}=\sum_{k=1}^{K}
\mathbf{h}_{kl}b_{k}s_k + \mathbf{n}_{l},
\end{align}
where $b_k$ is the transmit coefficient at wireless device $k$ with $\mathbb{E}\left\{|b_{k}s_k|^2\right\}=|b_{k}|^2\le P_k$, where $P_k$ is the maximum transmit power of wireless device $k$, and $\mathbf{n}_{l}\sim \mathcal{N}_{\mathbb{C}}\left(\mathbf{0}, \delta^2\mathbf{I}_N\right)$ is the noise vector at AP $l$. 

AP $l$ can send $\mathbf{y}_{l}$ and its local channel estimates to the CPU for centrally recovering the arithmetic mean function, or 
 partially compute the arithmetic mean function and share less information with the CPU. Therefore, there is a tradeoff between  better recovery of the arithmetic mean function and  fronthaul overhead. In this paper, we consider three levels of AP cooperation and develop optimal AirComp strategies for them respectively.

\subsection{{Fully Centralized Processing (Level 3)}}
At the highest level of AP cooperation, the CPU centrally performs channel estimation and recovers the arithmetic mean function. To this end, $\tau_c NL$ complex scalars consisting of $\tau_p NL$ complex scalars for pilot signals $\{\mathbf{Y}_{l}^{\text{pilot}}:l=1,\dots,L\}$ and $(\tau_c-\tau_p)NL$ complex scalars for data signals $\{\mathbf{y}_{l}:l=1,\dots,L\}$ are sent to the CPU via fronthaul links per coherence block.
Moreover, $KLN^2/2$ complex scalars for channel statistics $\{\mathbf{R}_{kl}:k=1,\dots,K, l=1,\dots,L\}$ are needed at the CPU for channel estimation and combing vector design \cite{bjornson2019making}. The fronthaul signaling is summarized in Table \ref{tab1}.

\begin{table*}[!ht]
\begin{center}
\caption{Number of Complex Scalars to Send via the Fronthaul Links.}
\label{tab1}
\begin{tabular}{| c | c | c | c |}
\hline
  & Each coherence time (APs $\to$ CPU) & Each coherence time (CPU $\to$ APs) &  Statistical parameters \\
\hline
\hline
Level 3 & $\tau_{c}NL$ & $K$ & $KLN^{2}/2$  \\
\hline
Level 2 & $\left(\tau_{c}-\tau_{p}\right)L$ & $-$ & $KL + \left(L + KL^{2}\right)/2$ \\
\hline
Level 1 & $\left(\tau_{c}-\tau_{p}\right)L$ & $-$ & $-$ \\
\hline
\end{tabular}
\end{center}
\end{table*}

After stacking the received data signals, the CPU obtains
\begin{align}
\mathbf{y}=\sum_{k=1}^{K}
\mathbf{h}_{k}b_{k}s_k + \mathbf{n},
\end{align}
where $\mathbf{y}=\left[\mathbf{y}_{1}^{T},\dots,\mathbf{y}_{L}^{T}\right]^{T}\in\mathbb{C}^{LN}$, $\mathbf{h}_{k}=\left[\mathbf{h}_{k1}^{T},\dots,\mathbf{h}_{kL}^{T}\right]^{T}\in\mathbb{C}^{LN}$, and $\mathbf{n} = \left[\mathbf{n}_{1}^{T},\dots,\mathbf{n}_{L}^{T}\right]^{T}\in\mathbb{C}^{LN}$. After channel estimation, the CPU forms the stacked channel estimate $\hat{\mathbf{h}}_{k}=\left[\hat{\mathbf{h}}_{k1}^{T},\dots,\hat{\mathbf{h}}_{kL}^{T}\right]^{T}\in\mathbb{C}^{LN}$ with estimate error $\tilde{\mathbf{h}}_{k}\sim \mathcal{N}_{\mathbb{C}}\left(\mathbf{0}, \mathbf{C}_{k}\right)$, where $\mathbf{C}_{k}=\text{diag}\left(\mathbf{C}_{k1},\dots,\mathbf{C}_{kL}\right)$.

The combing vector $\mathbf{v}\in\mathbb{C}^{LN}$ is designed to recover the arithmetic mean function as 
\begin{align}
\hat{f}_{(3)}=
\frac{\mathbf{v}^{H}\mathbf{y}}{K}=
\frac{1}{K}\left(\sum_{k=1}^{K}
{\mathbf{v}^{H}}\mathbf{h}_{k}b_{k}s_k + {\mathbf{v}^{H}}\mathbf{n}\right).
\end{align}
 The distortion between $\hat{f}_{(3)}$ and $f$ is evaluated using the MSE, which is given by
\begin{align}
\label{MSE}
&\mathbb{E}\left\{\left.\left|f-\hat{f}_{(3)}\right|^2  \right| \left\{\hat{\mathbf{h}}_{k}\right\}\right\} \nonumber \\
&=
\frac{1}{K^2}
\mathbb{E}\left.\left\{\left|\sum_{k=1}^{K}
\left(\mathbf{v}^{H}\mathbf{h}_{k}b_{k}-1\right)s_k + {\mathbf{v}^{H}}\mathbf{n}\right|^2 \right| \left\{\hat{\mathbf{h}}_{k}\right\}\right\} \nonumber \\
&=\frac{1}{K^2}
\left( \sum_{k=1}^{K}\left(
\left|\mathbf{v}^{H}\hat{\mathbf{h}}_{k}b_{k}-1\right|^{2} + |b_k|^{2}\mathbf{v}^{H}\mathbf{C}_{k}\mathbf{v} \right)
+ \delta^{2}\|\mathbf{v}\|^{2}\right).
\end{align}

 We aim to minimize the MSE by optimizing the transmit coefficients  $\{b_{k}: k=1,\dots,K\}$ and the combing vector $\mathbf{v}$. Accordingly, the optimization problem is formulated as
\begin{align}
\mathop{\min_{\{b_{k}\}, \mathbf{v}}}~& \mathbb{E}\left\{\left.\left|f-\hat{f}_{(3)}\right|^2  \right| \left\{\hat{\mathbf{h}}_{k}\right\}\right\}  \label{P0}   \\
      s.t.~          
      & |b_{k}|^2\le P_k, k=1,\dots,K. \tag{\ref{P0}{a}}  
\end{align}
The optimization problem is non-convex due to the coupling of $\{b_{k}\}$ and $\mathbf{v}$.  To this end, we propose to alternatively optimize $\{b_{k}\}$ and $\mathbf{v}$ as follows.

\subsubsection{Optimization of $\{b_{k}\}$} With a given $\mathbf{v}$, the subproblem of transmit coefficient optimization (TCO) is formulated as 
\begin{align}
\mathop{\min_{\{b_{k}\}}}~& \sum_{k=1}^{K}\left(
\left|\mathbf{v}^{H}\hat{\mathbf{h}}_{k}b_{k}-1\right|^{2} + |b_k|^{2}\mathbf{v}^{H}\mathbf{C}_{k}\mathbf{v} \right)  \label{P1}   \\
      s.t.~          
      & |b_{k}|^2\le P_k, k=1,\dots,K. \tag{\ref{P1}{a}}  
\end{align}
Since $b_{1}, \dots, b_{K}$ are decoupled, (\ref{P1}) can be further decomposed into $K$ subproblems. The $k^{\text{th}}$ subproblem is expressed as 
\begin{align}
\mathop{\min_{b_k}}~& 
\left|\mathbf{v}^{H}\hat{\mathbf{h}}_{k}b_{k}-1\right|^{2} + |b_k|^{2}\mathbf{v}^{H}\mathbf{C}_{k}\mathbf{v}  \label{P2}   \\
      s.t.~          
      & |b_{k}|^2\le P_k. \tag{\ref{P2}{a}}  
\end{align}

Attaching a Lagrange multiplier $\mu_{k}\ge 0$ to the constraint in (\ref{P2}a), we obtain the following Lagrangian function 
\begin{align}
\mathcal{L}(b_k)= \left|\mathbf{v}^{H}\hat{\mathbf{h}}_{k}b_{k}-1\right|^{2} + |b_k|^{2}\mathbf{v}^{H}\mathbf{C}_{k}\mathbf{v} + \mu_{k}\left(|b_{k}|^2-P_k\right).
\end{align}
The KKT conditions are given by
\begin{align}
\left|\mathbf{v}^{H}\hat{\mathbf{h}}_{k}\right|^{2}b_{k} - \hat{\mathbf{h}}_{k}^{H}\mathbf{v}+ b_{k}\left(\mathbf{v}^{H}\mathbf{C}_{k}\mathbf{v} + \mu_{k}\right)&=0, \label{KKT1}  \\
 \mu_{k}\left(|b_{k}|^2-P_k\right)& = 0, \label{KKT2} \\
 |b_{k}|^2 &\le P_k. 
\end{align}
The first-order optimality condition in (\ref{KKT1}) yields 
\begin{align}
\label{bk}
b_k^{\text{opt}}=\frac{\hat{\mathbf{h}}_{k}^{H}\mathbf{v}}{\left|\mathbf{v}^{H}\hat{\mathbf{h}}_{k}\right|^{2}+\mathbf{v}^{H}\mathbf{C}_{k}\mathbf{v} + \mu_{k}}.
\end{align}
 Plugging $b_k^{\text{opt}}$ into the complementary slackness condition in (\ref{KKT2}), we have
\begin{align}
\label{muk}
\mu_{k}^{\text{opt}}=\max\left(0, \frac{\left|\mathbf{v}^{H}\hat{\mathbf{h}}_{k}\right|}{\sqrt{P_k}} - \left|\mathbf{v}^{H}\hat{\mathbf{h}}_{k}\right|^{2} - \mathbf{v}^{H}\mathbf{C}_{k}\mathbf{v}\right).
\end{align}

\subsubsection{Optimization of $\mathbf{v}$}
The unconstrained optimization problem with regard to $\mathbf{v}$ is formulated as
\begin{align}
\label{eq:v}
\mathop{\min_{\mathbf{v}}}~& 
\sum_{k=1}^{K}\left(
\left|\mathbf{v}^{H}\hat{\mathbf{h}}_{k}b_{k}-1\right|^{2} + |b_k|^{2}\mathbf{v}^{H}\mathbf{C}_{k}\mathbf{v}\right) + \delta^{2}\|\mathbf{v}\|^{2}  
\end{align}
By setting the first derivative of the objective function in (\ref{eq:v}) with respective to $\mathbf{v}$ to zero, the optimal solution is obtained as
\begin{align}{\color{black}
\mathbf{v}^{\text{opt}}=\left(\sum_{k=1}^{K}
|b_k|^{2}\left(\hat{\mathbf{h}}_{k}\hat{\mathbf{h}}_{k}^{H}
 + \mathbf{C}_{k}\right)
 + \delta^{2}\mathbf{I}_{LN}
\right)^{-1}\sum_{k=1}^{K}b_k\hat{\mathbf{h}}_{k}.}
\end{align}

The alternating optimization is performed by iteratively updating $\{b_k^{\text{opt}}\}$ and $\mathbf{v}^{\text{opt}}$ until convergence. The convergence has been observed in our simulations, but has not been shown due to limited space. Finally, the well designed $\{b_k^{\text{opt}}\}$ need to be passed to the APs and then fed back to the wireless devices.

\subsection{{Local Processing \& Large-Scale Fading Decoding (Level 2)}}
At Level 2, the APs perform channel estimate locally and send the local estimates of the arithmetic mean function to the CPU.
 Let $\mathbf{v}_{l}\in \mathbb{C}^{N}$ denote the local combing vector at AP $l$. The local estimate of the arithmetic mean function at AP $l$ is given by
\begin{align}
\hat{f}_{l}=\frac{\mathbf{v}_{l}^{H}\mathbf{y}_{l}}{K}=
\frac{1}{K}\left(\sum_{k=1}^{K}
\mathbf{v}_{l}^{H}\mathbf{h}_{kl}b_{k}s_k + \mathbf{v}_{l}^{H}\mathbf{n}_{l}\right).
\end{align}
We design $\mathbf{v}_{l}$ to minimize the following MSE
\begin{align}
\label{eq:localMSE}
&\mathbb{E}\left\{\left.\left|f-\hat{f}_{l}\right|^2  \right| \left\{\hat{\mathbf{h}}_{kl}\right\}\right\} \nonumber \\
&=
\frac{1}{K^2}
\mathbb{E}\left.\left\{\left|\sum_{k=1}^{K}
\left(\mathbf{v}_{l}^{H}\mathbf{h}_{kl}b_{k}-1\right)s_k + {\mathbf{v}_{l}^{H}}\mathbf{n}_{l}\right|^2 \right|\left\{\hat{\mathbf{h}}_{kl}\right\}\right\} \nonumber \\
&=\frac{1}{K^2}
\left( \sum_{k=1}^{K}\left(
\left|\mathbf{v}_{l}^{H}\hat{\mathbf{h}}_{kl}b_{k}\!-\!1\right|^{2} \!+\! |b_k|^{2}\mathbf{v}_{l}^{H}\mathbf{C}_{kl}\mathbf{v}_{l} \right)
\!+\! \delta^{2}\|\mathbf{v}_{l}\|^{2}\right).
\end{align}
 By equating the first derivative of (\ref{eq:localMSE}) with regard to $\mathbf{v}_{l}$ to zero, the optimal $\mathbf{v}_{l}$ is obtained as
\begin{align}
\mathbf{v}_{l}^{\text{opt}}=\left(\sum_{k=1}^{K}
|b_k|^{2}\left(\hat{\mathbf{h}}_{kl}\hat{\mathbf{h}}_{kl}^{H}
 + \mathbf{C}_{kl}\right)
 + \delta^{2}\mathbf{I}_{N}
\right)^{-1}\sum_{k=1}^{K}b_k\hat{\mathbf{h}}_{kl},
\end{align}
which yields the optimal local estimate of the arithmetic mean function as
\begin{align}
\hat{f}_{l}^{\text{opt}}=
\frac{1}{K}\left(\sum_{k=1}^{K}
\left(\mathbf{v}_{l}^{\text{opt}}\right)^{H}\mathbf{h}_{kl}b_{k}s_k + \left(\mathbf{v}_{l}^{\text{opt}}\right)^{H}\mathbf{n}_{l}\right).
\end{align}
The optimal local estimates $\left\{\hat{f}^{\text{opt}}_{l}:l=1,\dots,L\right\}$ are then sent to the CPU for final recovery of the arithmetic mean function. The CPU combines the optimal local estimates using coefficients $\left\{a_{l}:l=1,\dots,L\right\}$ and obtains
\begin{align}
\hat{f}_{(2)}&=\sum_{l=1}^{L}a^{*}_{l}\hat{f}_{l}^{\text{opt}} \nonumber \\
&=
\frac{1}{K}
\sum_{l=1}^{L}a^{*}_l\left(\sum_{k=1}^{K}
\left(\mathbf{v}_{l}^{\text{opt}}\right)^{H}\mathbf{h}_{kl}b_{k}s_k + \left(\mathbf{v}_{l}^{\text{opt}}\right)^{H}\mathbf{n}_{l}\right) \nonumber \\
&=\frac{1}{K}
\left(\sum_{k=1}^{K}
\mathbf{a}^{H}\mathbf{g}_{k}b_{k}s_k + \mathbf{a}^{H}\mathbf{m}\right),
\end{align}
where $\mathbf{a}\!=[a_1,\dots,a_{L}]^{T}$, $\mathbf{g}_{k}\!=[(\mathbf{v}_{1}^{\text{opt}})^{H}\mathbf{h}_{k1},\dots,(\mathbf{v}_{L}^{\text{opt}})^{H}\mathbf{h}_{kL}]^{T}$,
and 
$\mathbf{m}=\left[\left(\mathbf{v}_{1}^{\text{opt}}\right)^{H}\mathbf{n}_{1},\dots,\left(\mathbf{v}_{L}^{\text{opt}}\right)^{H}\mathbf{n}_{L}\right]^{T}$.
Ideally, given $\{\mathbf{g}_k:k=1,\dots,K\}$ and $\{\mathbf{v}_{l}^{\text{opt}}:l=1,\dots,L\}$, the combing coefficient vector $\mathbf{a}$ is designed to minimize the following MSE
\begin{align}
\label{eq:MSE2}
&\mathbb{E}\left\{\left.\left|f-\hat{f}_{(2)}\right|^2  \right| \left\{\mathbf{g}_{k}\right\}, \left\{\mathbf{v}_{l}^{\text{opt}}\right\}\right\} \nonumber \\
&\ \ \ \ \ \ \ \ \ \ = \frac{1}{K^2}
\left( \sum_{k=1}^{K}
\left|\mathbf{a}^{H}{\mathbf{g}}_{k}b_{k}-1\right|^{2} 
+ \delta^{2}\mathbf{a}^{H}\mathbf{D}\mathbf{a}\right),
\end{align}
where $\mathbf{D}=\text{diag}\left(\|\mathbf{v}_{1}^{\text{opt}}\|^{2}, \dots, \|\mathbf{v}_{L}^{\text{opt}}\|^{2}\right)$. Then we have
\begin{align}
\label{eq:a}
\mathbf{a}=\left(\sum_{k=1}^{K}
|b_k|^{2}\mathbf{g}_{k}\mathbf{g}_{k}^{H}
 + \delta^{2}\mathbf{D}
\right)^{-1}\sum_{k=1}^{K}b_k{\mathbf{g}}_{k}.
\end{align}
However, the effective channels $\{\mathbf{g}_{k}\}$ and the optimal local combing vectors $\{\mathbf{v}_{l}^{\text{opt}}\}$ are unknown at the CPU. Instead, we use the channel statistics $\left\{\mathbb{E}\left\{\mathbf{g}_{k}\mathbf{g}_{k}^{H}\right\}:k=1,\dots,K\right\}$, $\mathbb{E}\left\{\mathbf{D}\right\}$, and 
$\left\{\mathbb{E}\left\{\mathbf{g}_{k}\right\}:k=1,\dots,K\right\}$ to recover the arithmetic mean
function. This approach is referred to as large-scale fading decoding~\cite{nayebi2016performance, bjornson2019making}.
The achievable optimal $\mathbf{a}$ is obtained by approximating (\ref{eq:a}) as
{\color{black}
\begin{align}
\mathbf{a}^{\text{opt}}=\left(\sum_{k=1}^{K}
|b_k|^{2}\mathbb{E}\left\{\mathbf{g}_{k}\mathbf{g}_{k}^{H}\right\}
 + \delta^{2}\mathbb{E}\left\{\mathbf{D}\right\}
\right)^{-1}\sum_{k=1}^{K}b_k\mathbb{E}\left\{\mathbf{g}_{k}\right\}.
\end{align}}

In summary, $(\tau_c - \tau_p)L$ complex scalars 
for $\left\{\hat{f}^{\text{opt}}_{l}\right\}$ are passed to the CPU per coherence block. As for the channel statistics, $L/2$  complex scalars for the real-valued diagonal matrix $\mathbb{E}\left\{\mathbf{D}\right\}$, $KL$ complex scalars for $\{\mathbb{E}\left\{\mathbf{g}_{k}\right\}\}$, and 
$KL^2/2$ complex scalars for Hermitian complex matrices
$\left\{\mathbb{E}\left\{\mathbf{g}_{k}\mathbf{g}_{k}^{H}\right\}\right\}$ are required to be sent to the CPU via fronthaul links, which are listed in Table \ref{tab1}.

Following similar steps for the optimization problem in (\ref{P1}),
the $b_{k}$ that minimizes the MSE in (\ref{eq:MSE2}) is given by 
\begin{align} 
b_k&=\frac{\mathbf{g}_{k}^{H}\mathbf{a}}{\left|\mathbf{a}^{H}\mathbf{g}_{k}\right|^{2}+ \mu_{k}}, \\
\mu_{k}&=\max\left(0, \frac{\left|\mathbf{a}^{H}\mathbf{g}_{k}\right|}{\sqrt{P_k}} - \left|\mathbf{a}^{H}\mathbf{g}_{k}\right|^{2} \right).
\end{align} 
Then the CPU utilizes channel statistics to obtain the following approximations: 
\begin{align} 
b_k&=\frac{\mathbb{E}\left\{\mathbf{g}_{k}^{H}\right\}\mathbf{a}}{\left|\mathbf{a}^{H}\mathbb{E}\left\{\mathbf{g}_{k}\right\}\right|^{2}+ \mu_{k}}, \\
\mu_{k}&=\max\left(0, \frac{\left|\mathbf{a}^{H}\mathbb{E}\left\{\mathbf{g}_{k}\right\}\right|}{\sqrt{P_k}} - \left|\mathbf{a}^{H}\mathbb{E}\left\{\mathbf{g}_{k}\right\}\right|^{2} \right).
\end{align} 
Nevertheless, the iterative updates of $\{b_k\}$ require new estimates of channel statistics $\{\mathbb{E}\left\{\mathbf{g}_{k}\right\}\}$ ($b_k$ is included in $\mathbb{E}\left\{\mathbf{g}_{k}\right\}$), which is computationally intensive and fronthaul-unfriendly. Hence, at Level 2, $b_{k}$ is simply set to $b_{k}=\sqrt{P_{k}}, k=1,\dots,K$.

\subsection{{Local Processing \& Simple Centralized Decoding (Level 1)}}
At Level 1, no channel statistics is required at the CPU,  but only the local estimates of the arithmetic mean function. The number of complex scalars sent to the CPU per coherence block is the same as Level 2, as listed in Table \ref{tab1}.
After receiving the 
optimal local estimates $\left\{\hat{f}_{l}^{\text{opt}}\right\}$, the CPU simply takes the average, i.e., $a_l=\frac{1}{L}, l=1,\dots,L$,  and recovers the arithmetic mean function as
\begin{align}
\hat{f}_{(1)}&=\frac{1}{L}\sum_{l=1}^{L}\hat{f}_{l}^{\text{opt}}.
\end{align} 
The transmit coefficients at the wireless devices are simply set to $b_{k}=\sqrt{P_{k}}, k=1,\dots,K$.

\section{Asymptotic Analysis}
\label{sec:Asymptotic Analysis}
In this section, will show that in the presence of channel estimation error, the MSE will not be zero even if the wireless devices' transmit powers approach infinity. The asymptotic analysis is based on Level 3, and thus the conclusion applies to all cooperation levels.
When $P_k \to \infty, k= 1,\dots, K$, from (\ref{muk}) we have $\mu_k^{\text{opt}}=0$ and $b_k^{\text{opt}} = \frac{\hat{\mathbf{h}}_{k}^{H}\mathbf{v}}{\left|\mathbf{v}^{H}\hat{\mathbf{h}}_{k}\right|^{2}+\mathbf{v}^{H}\mathbf{C}_{k}\mathbf{v}}$. Plugging $b_k^{\text{opt}}$ into (\ref{MSE}), the MSE can be rewritten as
\begin{align}
&\mathbb{E}\left\{\left.\left|f-\hat{f}_{(3)}\right|^2  \right| \left\{\hat{\mathbf{h}}_{k}\right\}\right\}  \nonumber \\
&\ \ \ \ \ =\frac{1}{K^2}
\left( \sum_{k=1}^{K} \frac{\mathbf{v}^{H}\mathbf{C}_{k}\mathbf{v}}{\left|\mathbf{v}^{H}\hat{\mathbf{h}}_{k}\right|^{2} + \mathbf{v}^{H}\mathbf{C}_{k}\mathbf{v}}
+ \delta^{2}\|\mathbf{v}\|^{2}\right) \nonumber \\
&\ \ \ \ \ =\frac{1}{K^2}
\left( \sum_{k=1}^{K} \frac{1}{\frac{\mathbf{v}^{H}\hat{\mathbf{h}}_{k}\hat{\mathbf{h}}_{k}^{H}\mathbf{v}}{\mathbf{v}^{H}\mathbf{C}_{k}\mathbf{v}} + 1}
+ \delta^{2}\|\mathbf{v}\|^{2}\right) \nonumber \\
&\ \ \ \ \ \ge \frac{1}{K^2}
\left( \sum_{k=1}^{K} \frac{1}{\hat{\mathbf{h}}_{k}^{H}\mathbf{C}_{k}^{-H}\hat{\mathbf{h}}_{k} + 1}
+ \delta^{2}\|\mathbf{v}\|^{2}\right),
\end{align}
{\color{black}where the inequality holds because $\frac{\mathbf{v}^{H}\hat{\mathbf{h}}_{k}\hat{\mathbf{h}}_{k}^{H}\mathbf{v}}{\mathbf{v}^{H}\mathbf{C}_{k}\mathbf{v}}$ is a generalized Rayleigh quotient with respect to $\mathbf{v}$.} Hence, if $\mathbf{C}_{k}\neq \mathbf{0}, k= 1,\dots, K$, the MSE will not be zero even if $\|\mathbf{v}\|^{2}\to 0$.

\section{Numerical Results}
In this section, we present the MSE performance achieved by Cell-free mMIMO systems under different cooperation levels, as well as by Cellular mMIMO. We consider a  $1\times 1$ km simulation area and adopt the wrap-around topology to avoid unnatural boundaries. As for the Cellular mMIMO, a single base station equipped with 144 antennas is located in the center of the area. Since antenna arrays are collocated in Cellular mMIMO, fully centralized processing is employed to recover the arithmetic mean function.
The APs of the Cell-free  mMIMO network are deployed on a square grid. The total number of antennas in the Cell-free  mMIMO network is set to 144 for fair comparison.
There are $K=20$ wireless devices uniformly distributed in the considered area.
We adopt a random pilot assignment strategy and
 the uplink pilot powers are set to $p_k=20$ dBm, $k=1,\dots,K$. We assume a carrier frequency of 2 GHz, a bandwidth of 20 MHz and 
a noise power of $\delta^2=-96$ dBm.

We use the same propagation model for Cell-free  mMIMO and Cellular mMIMO for fair comparison.
Following the 3GPP Urban Microcell model~\cite{3gpp2010further}, the large-scale fading is given by 
\begin{align}
\beta_{kl}[\text{dB}]= \beta_{0} - 10\alpha \text{log}_{10}\left(\frac{d_{kl}}{d_0}\right) + S_{kl},
\end{align} 
where $d_{0}=1$~m is the reference distance, $\beta_{0}=-30.5$~dB is the large-scale path loss at the reference distance, $\alpha=3.67$ is the path-loss exponent, $d_{kl}$ is the distance between AP $l$ and wireless device $k$, and $S_{kl}\sim \mathcal{N}(0,4^2)$ is the shadow fading. The correlation between two shadow terms associated with AP $l$ is 
$\mathbb{E}\{S_{kl}S_{il}\}=4^2 2^{-x_{ki}/9 \ \text{m}}$, where $x_{ki}$ is the distance between wireless device $k$ and wireless device $i$~\cite{3gpp2010further}. The 
shadow terms associated with two different APs can be considered as uncorrelated due to the large inter-AP distance. All antenna arrays are assume to be uniform linear arrays with half-wavelength spacing. The widely used Gaussian local scattering model is adopted to model spatial channel correlation with $15^{\circ}$ angular standard deviation~\cite{bjornson2017massive}.

\subsection{Performance Evaluation of  AirComp}
\begin{figure}[t]
\centerline{\includegraphics[width=3.4in]{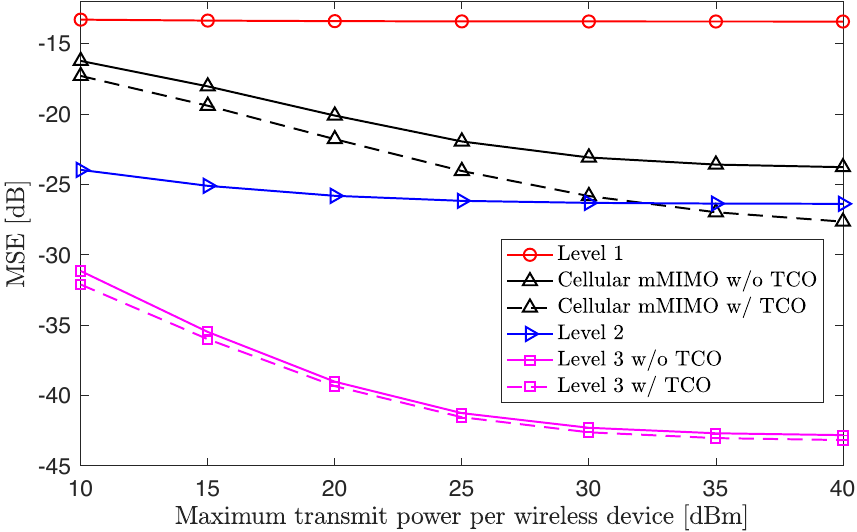}}
\caption{MSE versus the maximum transmit power per wireless device with {\color{black}$L=144$, $N=1$}, and {\color{black}$\tau_{p}=20$}.}
\label{fig:L144tau20}
\end{figure}

In Fig. \ref{fig:L144tau20}, we set $L=144,N=1$ and $\tau_p=20$, and show the MSE of AirComp versus the maximum transmit power budget per wireless device. 
As expected, a higher cooperation level leads to a smaller MSE.  In particular, the MSE of Level 3 without TCO is at least 9.7 dB lower than the Cellular mMIMO with TCO. Cellular mMIMO can only outperform  Level 2 at high power budgets and using TCO. Moreover, we observe that the benefit of using TCO in Cell-free mMIMO is not significant due to the uniform distribution of cooperating APs. This means we don't have to rely on TCO to achieve a small MSE.

\begin{figure}[t]
\centerline{\includegraphics[width=3.4in]{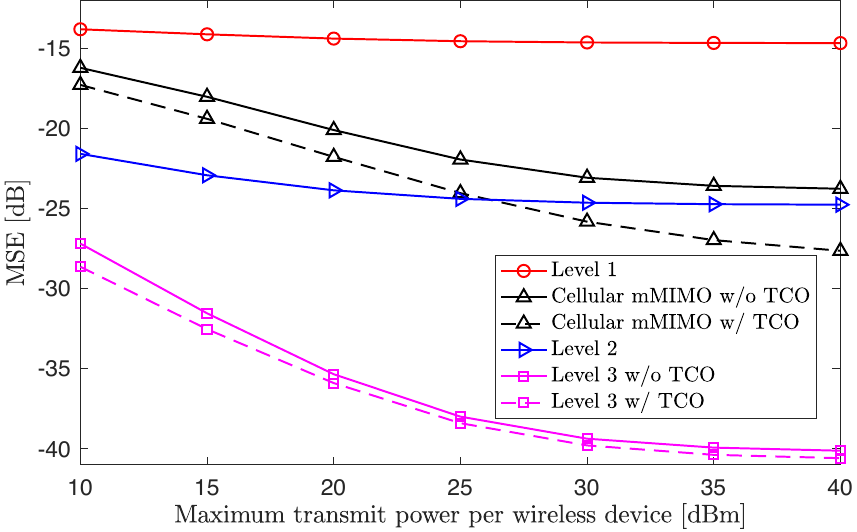}}
\caption{MSE versus the maximum transmit power per wireless device with {\color{black}$L=36$, $N=4$},  and {\color{black}$\tau_{p}=20$}.}
\label{fig:L36tau20}
\end{figure}

Fig. \ref{fig:L36tau20} presents the MSE versus the power budget per wireless device with $L=36,N=4$ and $\tau_p=20$. The curves show similar trends as in Fig. \ref{fig:L144tau20}, but the MSE increases at Level 2 and Level 3 due to the reduced macro diversity. However, Level 2 still outperforms Cellular mMIMO with regards to MSE at low power budgets. This has practical implications as wireless devices typically have limited power budgets. Compared to the results in Fig. \ref{fig:L144tau20}, Level 1 achieves a smaller MSE because each AP has a stronger local processing ability with more antennas.

\begin{figure}[t]
\centerline{\includegraphics[width=3.4in]{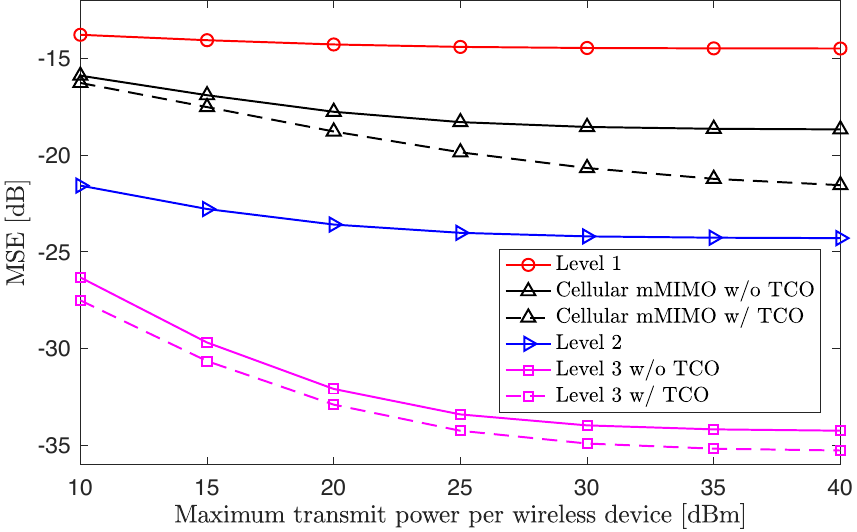}}
\caption{MSE versus the maximum transmit power per wireless device with {\color{black}$L=36$, $N=4$},  and {\color{black}$\tau_{p}=10$}.}
\label{fig:L36tau10}
\end{figure}

In Fig. \ref{fig:L36tau10}, we consider $L=36,N=4$ and $\tau_p=10$. Since $\tau_p< K$, there is pilot contamination. 
Compared to the results in Fig. \ref{fig:L36tau20}, both Cell-free mMIMO and Cellular mMIMO lose performance in terms of MSE due to the increased channel estimation 
errors. It is observed that Level 2 achieves a smaller MSE than Cellular mMIMO under various power budgets. Moreover, we can see that TCO brings an obvious performance gain in terms of MSE, demonstrating the benefit of TCO in the presence of strong pilot contamination.

 Furthermore, we observe that under various system parameter, for both Cell-free mMIMO and Cellular mMIMO systems, the MSE first decreases with the maximum transmit power per wireless device and then reaches a non-zero minimum value. This matches our asymptotic analysis in Section \ref{sec:Asymptotic Analysis}.

\subsection{Comparison of Fronthaul Signaling}
Although Level 3 achieves significantly lower MSE than Level 2 and 1, it may incur a large amount of fronthaul signaling exchange. To be specific, according to Table \ref{tab1}, Level 3 requires $\frac{\tau_c NL + K}{(\tau_c-\tau_p)L}$
times more fronthaul signaling than Level 2 and 1 in each coherence block. Hence, Level 3 will lead to a significantly higher fronthaul overhead than Level 2 and 1 if multi-antenna APs are used, i.e., $N>1$. This is because local estimates at Level 2 and 1 convert the $N$-dimensional data signals into scalars. Note that Level 2 and 1 may require more fronthaul signaling than Level 3 in situations where there are multiple groups of wireless devices and different groups have different AirComp tasks, which will be investigated in our extended journal version.

 \section{Conclusions}
\label{sec:conclusion}
In this paper, we studied the performance of AirComp in Cell-free mMIMO systems under three levels of AP cooperation. 
Simulation results showed that a higher level of AP cooperation achieves a lower MSE, but at the cost of more fronthaul signaling.
Fully centralized processing at Level 3 outperforms the Cellular mMIMO by a large margin, while partially distributed processing at Level 2 
can perform better than Cellular mMIMO for wireless devices with low power budgets. Since wireless devices are typically power-limited, Level 2 strikes a balance between MSE  and fronthaul signaling.
Moreover, we analytically and experimentally showed that a non-zero MSE is inevitable in the presence of channel estimation errors even if the wireless devices have unlimited power budgets. 
In the future, it will be of interest to investigate energy efficient AirComp in Cell-free mMIMO systems.

 \section{Acknowledgement}
This work was supported by the KTH Digital Future research center.




\end{document}